\documentclass[conference]{IEEEtran}

\IEEEoverridecommandlockouts
\usepackage{cite}
\usepackage{amsmath,amssymb,amsfonts}
\usepackage{algorithmic}
\usepackage{graphicx}
\usepackage{textcomp}
\usepackage{xcolor}
\usepackage{siunitx}
\usepackage{footnote}
\usepackage{tikz}
\usepackage{caption}
\usepackage{subcaption}
\usepackage{epstopdf}
\usetikzlibrary{arrows,shadows,positioning,calc}

\usepackage[hidelinks]{hyperref}

\usepackage{listings}
\definecolor{commentsColor}{rgb}{0.497495, 0.497587, 0.497464}
\definecolor{keywordsColor}{rgb}{0.000000, 0.000000, 0.635294}
\definecolor{stringColor}{rgb}{0.558215, 0.000000, 0.135316}

\def\BibTeX{{\rm B\kern-.05em{\sc i\kern-.025em b}\kern-.08em
    T\kern-.1667em\lower.7ex\hbox{E}\kern-.125emX}}

\begin{document}

\newcommand{\todo}[1]{{\color{red}TODO: #1}}

\title{SonicPACT: An Ultrasonic Ranging Method for the Private Automated Contact Tracing (PACT) Protocol\\
\thanks{DISTRIBUTION STATEMENT A. Approved for public release. Distribution is unlimited.\\ \\This material is based upon work supported by the Defense Advanced Research Projects Agency under Air Force Contract No. FA8702-15-D-0001 and the MIT–IBM Watson Artificial Intelligence Laboratory. Any opinions, findings, conclusions or recommendations expressed in this material are those of the author(s) and do not necessarily reflect the views of the Defense Advanced Research Projects Agency. \\ \\ R. L. Rivest received support from the Center for Science of Information (CSoI), an NSF Science and Technology Center, under grant agreement CCF-0939370. \\ \\ LC Ivers received funding from the Sullivan Family Foundation. \\ \\ Principal authors Meklenburg, Specter and Wentz each made equal contributions to research and analysis reported in this paper. }
}

\author{\IEEEauthorblockN{
	John Meklenburg\IEEEauthorrefmark{1}\IEEEauthorrefmark{2},
    Michael Specter\IEEEauthorrefmark{1}\IEEEauthorrefmark{3}, 
    Michael Wentz\IEEEauthorrefmark{1}\IEEEauthorrefmark{2}, \\
	Hari Balakrishnan\IEEEauthorrefmark{3},
	Anantha Chandrakasan\IEEEauthorrefmark{3},
	John Cohn\IEEEauthorrefmark{4},
    Gary Hatke\IEEEauthorrefmark{2},
	Louise Ivers\IEEEauthorrefmark{5},\\
    Ronald Rivest\IEEEauthorrefmark{3}, 
	Gerald Jay Sussman\IEEEauthorrefmark{3},
    Daniel Weitzner\IEEEauthorrefmark{3},\\
}
\IEEEauthorblockA{
	\IEEEauthorrefmark{1}Corresponding Authors,
    \IEEEauthorrefmark{2}MIT Lincoln Laboratory,
    \IEEEauthorrefmark{3}Massachusetts Institute of Technology,\\
	\IEEEauthorrefmark{4}MIT-IBM Watson AI Lab,
    \IEEEauthorrefmark{5}Massachusetts General Hospital
}
}

\maketitle

\begin{abstract}
Throughout the course of the COVID-19 pandemic, several countries have developed and released contact tracing and exposure notification smartphone applications (apps) to help slow the spread of the disease. To support such apps, Apple and Google have released Exposure Notification Application Programming Interfaces (APIs) to infer device (user) proximity using Bluetooth Low Energy (BLE) beacons. The Private Automated Contact Tracing (PACT) team has shown that accurately estimating the distance between devices using only BLE radio signals is challenging \cite{hatke}. This paper describes the design and implementation of the SonicPACT protocol to use near-ultrasonic signals on commodity iOS and Android smartphones to estimate distances using time-of-flight measurements. The protocol allows Android and iOS devices to inter-operate, augmenting and improving the current exposure notification APIs. Our initial experimental results are promising, suggesting that SonicPACT should be considered for implementation by Apple and Google.
\end{abstract}


\section{Introduction}

Since the onset of the COVID-19 pandemic there has been significant interest in the use of smartphone applications (apps) to facilitate contact discovery and tracing. Researchers have been pursuing methods using Bluetooth Low Energy (BLE) signals to estimate the duration and proximity of smartphones (and those carrying them) to one another. One such effort is the Private Automated Contact Tracing (PACT) Protocol~\cite{pact}, which is a decentralized approach modeled after Apple's ``Find My" protocol~\cite{findmy}. A smartphone using a PACT-based app periodically emits BLE chirps derived from a secret seed known only to that phone. All smartphones log their seeds, the times the seeds were active, and any BLE chirps received with timestamps. If a user is diagnosed as COVID-19 positive, a record of all BLE chirps sent by the user's phone can be voluntarily uploaded to a database through a trusted public health authority. Other smartphones frequently check against this database to see if they have received any BLE chirps designated from a COVID-19 positive individual, and if so, the phone's owner can be notified to take the appropriate action (for instance, self-isolate or seek COVID-19 testing).

In order for PACT and related efforts to be successful, a Too Close For Too Long (TCFTL) detector must be developed based on suitable signals (e.g., BLE) and associated measurements, such as the transmit power level, Received Signal Strength Indicator (RSSI), and received timestamp. There are three desired features of the TCFTL detector:
\begin{enumerate}
\item a high probability of detection when the two users (devices) are within a threshold distance (i.e., high recall),
\item low probability of false alarm (high precision), and
\item ability to be tuned based on guidelines from local health authorities (for example, based on an exposure of 15 or more minutes within 6 feet of an infected individual \cite{cdc}).
\end{enumerate}

Apple and Google ($A \vert G$) have released BLE-based exposure notification protocols and Application Programming Interfaces (APIs) for iOS and Android devices, a significant step forward for automated contact tracing efforts~\cite{applegoogle}. Unfortunately, the initial implementations of these protocols only log received BLE chirps when the device is awake, or no more often than every 5 minutes when the device is not awake. This limits the data available to a TCFTL detector, and it has been shown to have relatively poor performance (up to 50\% false alarms and less than 25\% of true contact events detected), with an effectiveness that varies across smartphones~\cite{hatke, trinitygaen, trinitybus, trinitytram, trinitycal}. Increasing the BLE sample rate and estimating the carriage state of each smartphone has been shown to help~\cite{hatke}, although may not be possible due to battery life and privacy constraints.

In this paper, we show that ultrasonic (US)\footnote{Ultrasonic frequencies are typically defined as those exceeding 20 kHz. In the context of commodity smartphones and this paper, we use ``ultrasonic" for frequencies around 20 kHz, but not necessarily exceeding that threshold.} range measurement using the speakers and microphones built-in to all smartphones is a promising concept that could improve the accuracy of BLE-based estimates significantly. By using a BLE protocol augmented with US ranging (BLE+US), smartphones exchange time-tagged inaudible acoustic pulses in the ultrasonic frequency range to measure the time-of-flight between them. Because the speed of sound is known, the devices can then jointly solve for the range between them. By improving the ranging accuracy, this technique can be used together with BLE-based protocols to reduce the false alarm and miss rates of exposure notification services. For example, a BLE-based approach may flag neighbors in an apartment building as positive contacts, when they are actually separated by physical barriers (ceilings or walls) because the 2.4 GHz BLE signals will propagate through such barriers. A BLE+US approach would identify these BLE contacts as false alarms and eliminate them, since US frequencies are more significantly attenuated by the same barriers.

This paper introduces {\bf SonicPACT}, an ultrasonic ranging protocol developed by the PACT team to augment a BLE-based TCFTL detector for contact tracing. SonicPACT has the following key features:

\begin{enumerate}

\item It can be implemented on both Android and iOS and allow interoperability between them.
\item It operates at inaudible near-ultrasonic frequencies to avoid disturbing users while measurements are made.
\item It uses BLE advertisements (rather than Wi-Fi, although it could also use Wi-Fi) to coordinate measurements and exchange information. This feature simplifies integration with BLE-based exposure notification protocols.
\item It uses pseudorandom noise (PN) waveforms generated by a random number generator with a seed based on the device's unique Bluetooth Universally Unique Identifier (UUID), allowing both devices to generate matched filters for the other's waveform without prior coordination.

\end{enumerate}

We have implemented SonicPACT as user-level apps on Android and iOS and have conducted several experiments with the initial implementation. Our results are promising:
\begin{enumerate}
    \item In experiments done placing phones at distances between 2 feet (60 cm) and 12 feet (3,6 meters) of each other in multiples of 2 feet, and considering a threshold of $\leq$ 6 feet as the critical distance at which to flag a ``contact'', we find that the missed detection rate indoors is between 0\% and 5.1\% with no false alarms. 
    \item Indoors, the estimated distances are within 1 foot (30 cm) of the true distance between 56.7\% and 70.9\% of the time; outdoors, due to the absence of significant multipath, estimates are within 1 foot between 82.9\% and 100\% of the time.
    \item These are proof-of-concept implementations with significant optimizations that can be done, but the low false alarm and miss rates compared to a BLE-only method when estimating if two devices are ``too close'' (within 6 feet or 1.8 meters) is an encouraging result for exposure notification and is therefore worth studying for implementation by $A \vert G$.
\end{enumerate}

Our goal in developing SonicPACT was to demonstrate that ultrasonic signaling, coupled with BLE technology widely available on $A \vert G$ APIs, can provide a more robust implementation of a TCFTL detector. Ultimately, a power-efficient and ubiquitous implementation of an ultrasonic ranging protocol may (will?) need to be integrated within the iOS and Android operating systems, and may therefore need to be developed by $A \vert G$ themselves. As such, SonicPACT as proposed here is not a turnkey app solution to contact discovery, but we view it as a promising technology and protocol to incorporate into Android and iOS. In addition to discussing our initial promising experimental results, we discuss at length the limitations of both our SonicPACT implementation and of ultrasonic ranging in general, along with directions for future work.



\subsection{Related Work}
\label{sec:related}


Several research groups have sought to use the acoustic channel on smartphones or mobile devices for communications \cite{lee, jiang, getreuer, iannucci2015room}, and indoor and outdoor localization \cite{cricket, priyantha2005, activebat, girod2001robust, peng, lazikrowe} with some success. Most prior acoustic indoor localization techniques rely on external hardware to serve as beacons or receivers. Ranging between smartphones without any external hardware is more challenging, since depending on the method employed, accurate transmit and receive timestamps may be required to compute time-of-flight of the US signal between devices. Obtaining the timestamps with low or predictable latency from a user-space application is not easy.

Peng et al.'s BeepBeep method \cite{peng} provides a ranging capability using only commercial off-the-shelf (COTS) hardware available on smartphones. BeepBeep ranging uses linear frequency modulated (LFM) chirp waveforms in the 1-6 kHz frequency range and a sample-counting technique to eliminate the latency and uncertainty introduced by time-stamping the transmit and receive pulses. The protocol uses Wi-Fi to exchange measurements between devices, enabling each device to compute the range to the other device. The BeepBeep paper reports that sub-centimeter ranging accuracy was achievable in many cases, and that performance could be achieved even in challenging indoor environments with multipath mitigation techniques. Fotouhi \cite{fotouhi2} extended this work by implementing the BeepBeep protocol on Android devices using spread-spectrum waveforms to differentiate between devices and maintain user privacy. 

In April 2020, Loh released the NOVID app, which was the first-available COVID-19 contact tracing app to use both Bluetooth and ultrasound for contact detection. Loh's laboratory tests of the app shows excellent ranging performance in a controlled setting \cite{novid}. At this time, the technical details of the NOVID implementation have not been made available to the public. 

\subsection{Security and Privacy Concerns}

Here we review the security and privacy trade-offs inherent in the use of SonicPACT, or any exposure notification app that uses acoustic measurement.

\textbf{Potential Privacy Risks:} Any acoustic measurement implementation will require the ability to process audio, which poses a risk to user privacy. At the outset, we suggest it is appropriate to assess the privacy risks of this protocol relative to the design and implementation of the current ($A \vert G$) service. This protocol ought to satisfy the core privacy and security commitments of the ($A \vert G$) BLE protocol, including 

\begin{itemize}
\item voluntary adoption, implemented by requiring affirmative opt-in to the service for all users
\item protection of user location privacy, including no collection of GPS or other location information
\item confidentiality protection for all users of the system, as between potential contacts, between those contacts and the public health authority, and as may be observed by third party adversaries
\end{itemize}

Requiring constant audio access also allows an adversary with control over the operating system or ranging app itself to covertly record potentially sensitive conversations, and normalizing the use of such systems may potentially lead to overreach by the developers involved.  We believe, however, that such risks can be mitigated through carefully designed technical and policy mechanisms to avoid leakage or overreach. 

First, we advocate that the protocol be implemented by the mobile operating system vendor (e.g., Apple or Google) at the OS level via a privileged app, or as a part of the audio processing path itself. 
This approach can limit the exposure of such data to trusted applications with OS-level privileges, \footnote{This could be the SEPolicy equivalent of \texttt{priv\_app} or \texttt{system\_app} for Android.}
which, beyond preventing leakage to apps that would already be capable of such spying, would also represent the same risk profile as currently-existing wake-word activation (``Hey Siri'' and ``OK Google'').

Furthermore, ultrasonic sensing does not require the audible range of the spectrum. Any frequency below, say, 18 kHz may (should) be filtered and removed by the operating system (or, if possible, at the hardware level), ensuring that the app is only capable of listening at near-US frequencies. 

Finally, we would expect that there would be little need to store any information about prior interactions after a range as been estimated. Indeed, our proof of concept implementations discard audio samples after processing, and such information is never intentionally stored on disk. We recommend the same no-storage policy for a production implementation.

\textbf{Resolving Weaknesses in the  ($A \vert G$) Protocol:} 
A weakness of the ($A \vert G$) protocol is that Bluetooth packets may be heard, recorded, and rebroadcast later (a \emph{replay attack}) or somewhere else (a \emph{relay attack}). An interactive protocol, such as the SonicPACT protocol presented in this paper, can be augmented to avoid this weakness. Such protection may be obtained
by blending SonicPACT with a standard Diffie-Hellman key-agreement
protocol. 

Here, each device's unique identifier key can serve a dual purpose as a freshly-generated
Diffie-Hellman public key (as is done in the Apple Find My protocol~\cite{findmy}).
A pair of devices can therefore create an agreed upon shared secret
key $K$, which can then be used to 
seed the generation of the waveforms used in the protocol, and to generate message authentication
codes (MACs) for additional confirmation messages after the fact. If the parties use  distinct keys 
for each session bounded by some fixed time period, then they also have ``freshness'' guarantees since the cryptographic key $K$ will inherently depend on the time period.  


\section{Laptop Experiment}
\label{sec:laptop}

Before starting on Android and iOS devices, we conducted experiments using laptops as smartphone surrogates to demonstrate that accurate US ranging is possible without specialized hardware, and to understand what the performance of smartphone ranging performance might be. We used the built-in laptop speakers and microphones to exchange US pulses and processed the signals on both laptops using MATLAB software to compute the range between the devices.

We used LFM chirp waveforms ranging with the parameters listed in Table \ref{tab1}. Laptop A (a Dell Precision 5540) transmitted up-chirps (frequency swept from low to high) while Laptop B (a MacBook Pro, 2018 model, 13-inch display) transmitted down-chirps (frequency swept from high to low) to allow the processing code to easily differentiate between the two signals using a matched filter. We staggered the signal transmissions by 1 second to avoid interference. After data collection, we matched the signals recorded on both laptops against both transmit waveforms, yielding four signal comparisons between laptops A and B:

\begin{enumerate}
\item A's recording against A's transmit waveform
\item A's recording against B's transmit waveform
\item B's recording against A's transmit waveform
\item B's recording against B's transmit waveform
\end{enumerate}
Then, we computed the range between laptops using the sample-counting method described in \cite{peng}.

\begin{figure}[t]
\centerline{\includegraphics[width=\linewidth]{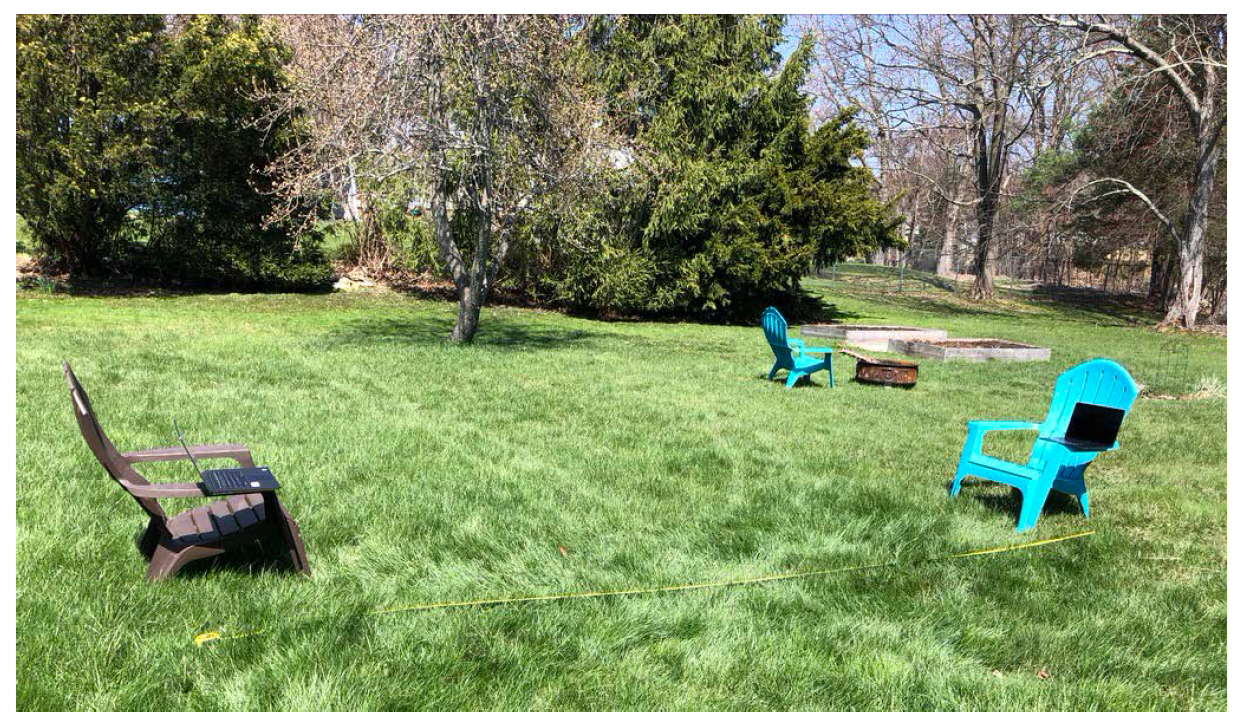}}
\caption{Laptop experiment, outdoor setup.}
\label{laptop_experiment}
\end{figure}

We conducted an outdoor experiment by setting the laptops on lawn chairs as shown in Figure \ref{laptop_experiment}. We varied the distance between laptops and measured the ground truth distance between them with a tape measure. We took US range measurements at each position by averaging around 10 pulse exchanges.

\begin{figure}[ht]
\centerline{\includegraphics[width=\linewidth]{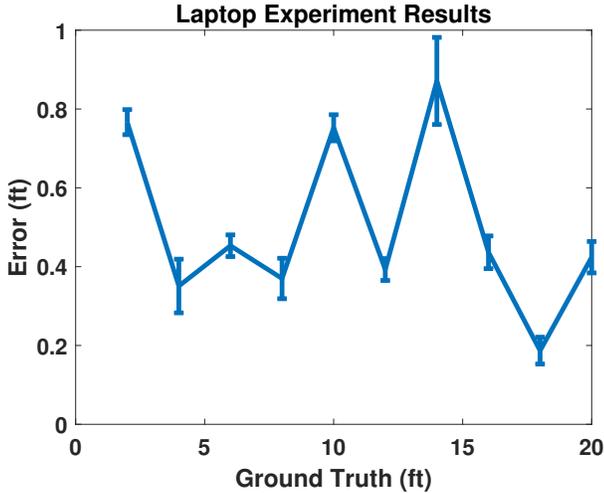}}
\caption{Laptop experiment results: range measurement vs truth. 1 ft = 30 cm.}
\label{fig1}
\end{figure}

The results (Figure \ref{fig1}) show that the measured range values were consistently within 1 foot (30 cm) of ground truth. Likely explanations for the apparent systematic error that manifests as a bias in Figure \ref{fig1} include multipath and non-colocated speakers and microphones on the laptops.

\begin{table}
\caption{Laptop Experiment Waveform Parameters}
\begin{center}
\begin{minipage}{6cm}
\begin{tabular}{lll}
\textbf{Parameter}        & \textbf{Value}           &      \\
Pulse Width               & 200 ms &      \\
Pulse Repetition Interval & 2 s                      &     \\
Center Frequency          & 20 kHz                   &      \\
Sampling Rate             & 48 kHz                   &      \\
Bandwidth                 & 2 kHz                    &      \\
Rise/Fall Time (to reduce audible clicks)           & 10 ms  &     
\end{tabular}
\end{minipage}
\label{tab1}
\end{center}
\end{table}

\section{System Design}
\label{sec:design}

Though BeepBeep served as a starting point for our work, the protocol has a few weaknesses that make it less functional for practical, widespread adoption. SonicPACT attempts to improve on this in the following ways: 

\begin{itemize}
  \item The audio signal processing uses LFM chirps that are not easily differentiable between devices and will be more difficult to scale to tens or hundreds of devices.
  \item BeepBeep's broadcasts are in audible frequency ranges
\end{itemize}

\subsection{Protocol}
\label{sec:protocol}

The intent of SonicPACT is to act in concert with BLE measurement estimates that indicate that the devices may be in range for a sufficient amount of time (as defined by public health authorities), following the ($A \vert G$) Exposure Notification framework. Each phone maintains a table of BLE UUIDs (or other identifiers) and the length of time for which they have been observed over a sliding window, for instance 15 minutes. The table is updated during each BLE scan period, and if packets are received from a device for greater than or equal to some threshold (which we anticipate would be lower than the threshold for contact duration), then SonicPACT ultrasonic waveforms are exchanged to more accurately estimate the proximity between phones.

SonicPACT must then perform the following tasks after BLE has initiated the ranging process:
\begin{enumerate}
    \item Generate the ranging waveforms, ensuring that the random samples are consistent between devices (drawn from the same random number generator using the same seed). 
    \item Transmit and receive audio, accounting for any delays by using a transmit-receive loopback measurement. 
    \item All devices must be able to detect waveforms from both themselves and nearby counterparts. \item Finally, each pair of nearby devices exchanges timing measurements over BLE to compute the range to the other device.
    \end{enumerate}



\begin{figure*}[]
\centerline{\includegraphics[width=\textwidth]{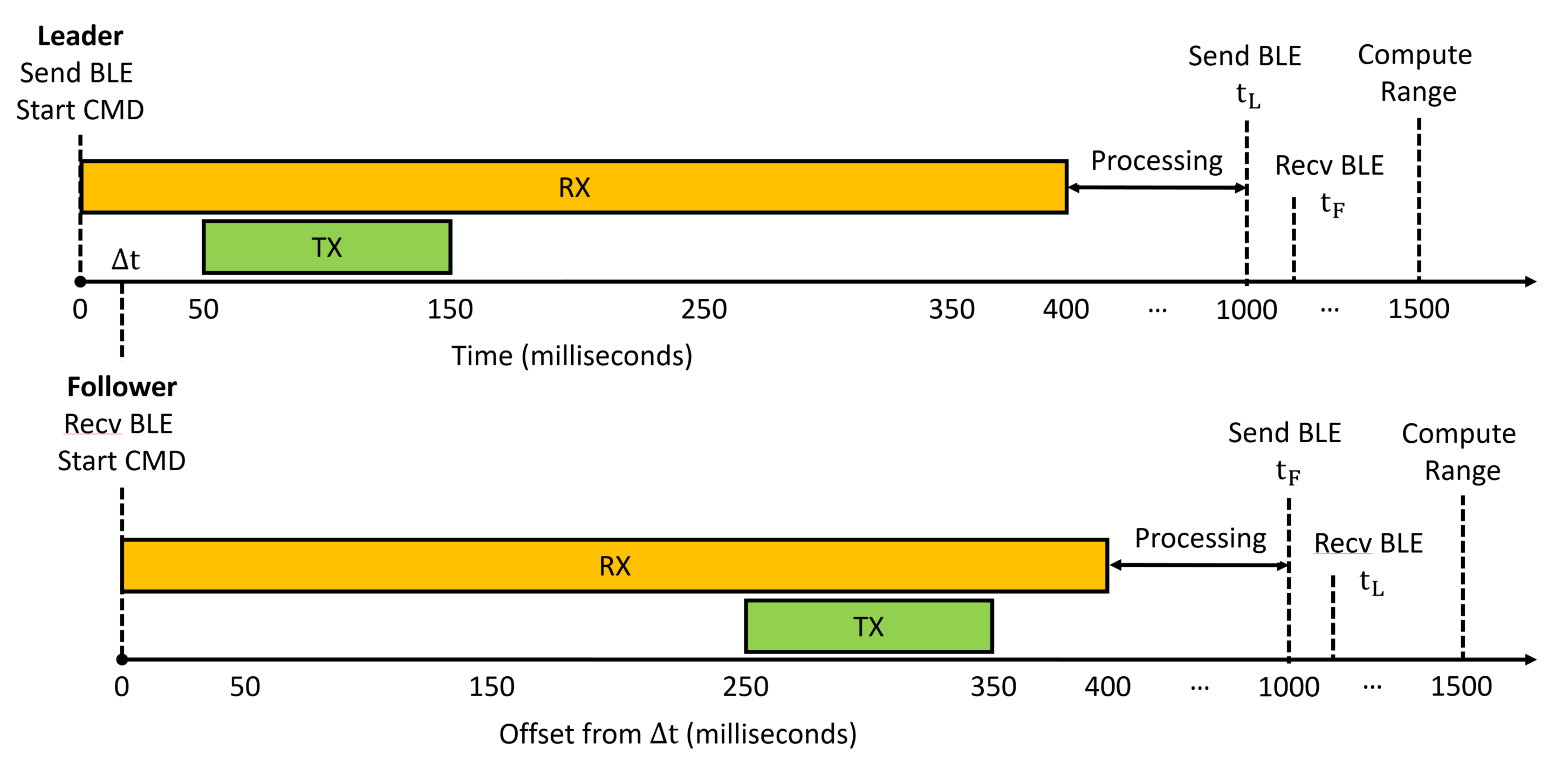}}
\caption{Prototype BLE/Ultrasonic Protocol.}
\label{protocol_orig}
\end{figure*}

The SonicPACT protocol is shown in Figure \ref{protocol_orig}. Each phone generates waveform samples for its own UUID and prepares to transmit that signal. To process a received waveform, a phone must know the UUID of the transmitter, which has been previously broadcast on the BLE channel. We describe here how a pair of phones can estimate their range, but this method can be extended to multiple devices in the same neighborhood as well.

The two phones select a leader and follower, so they know roughly when to expect the other’s waveform. In practice this could be chosen a number of ways. One  idea is to compute an integer hash from the UUIDs of both phones and always choose the lower integer to be the leader. The leader then starts the process by sending a start command via BLE. In our implementation, the leader sends its ultrasonic signal after a delay of 50 ms, and the follower sends its ultrasonic signal 250 ms after receiving the BLE start command. Both phones receive samples for a total of 400 ms, after which they perform matched filtering and detection for their own waveform and the other phone’s waveform. These are unoptimized and over-engineered parameters to demonstrate the feasibility of the underlying concepts.

\begin{figure}[]
\centerline{\includegraphics[width=\linewidth]{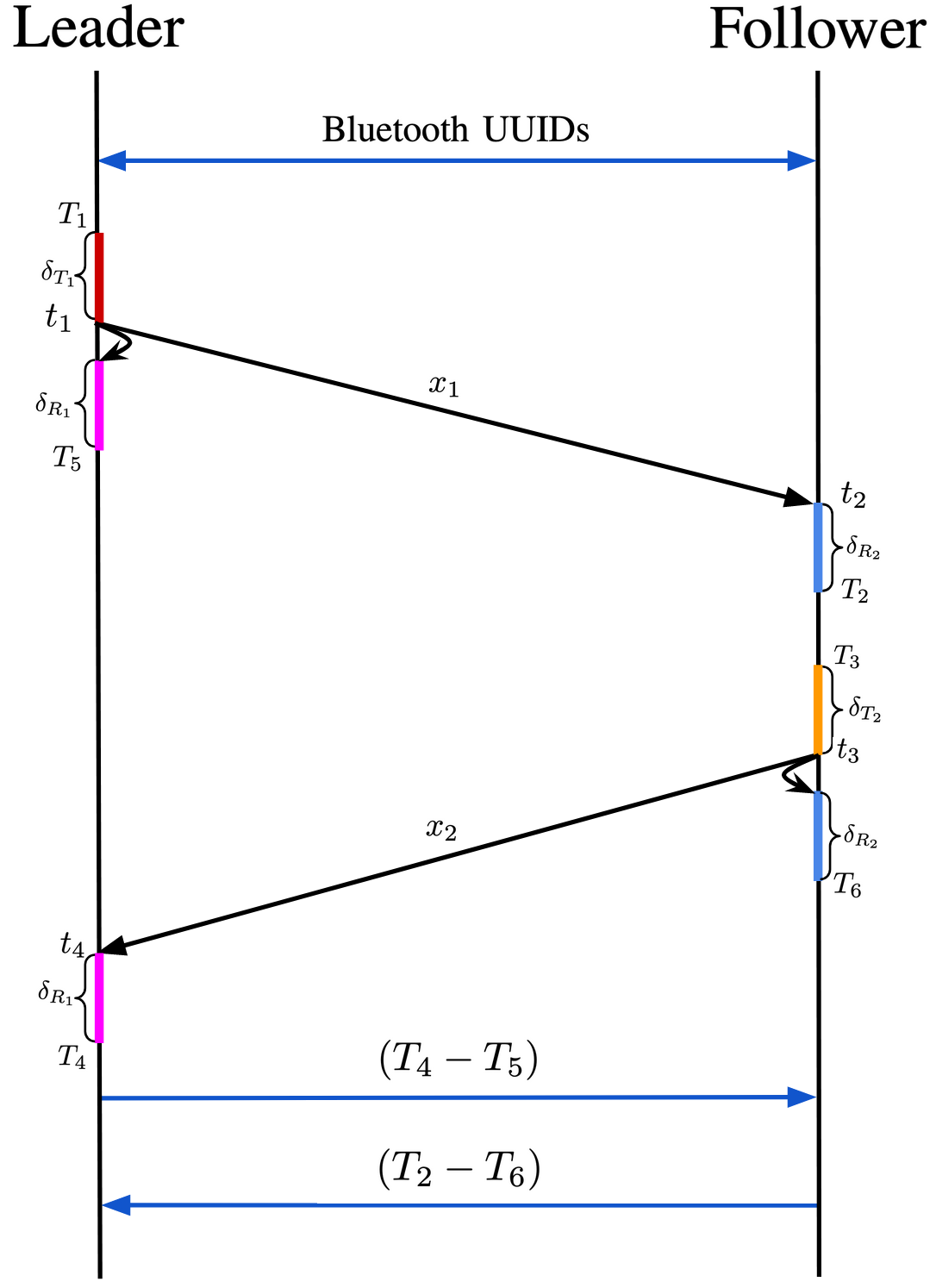}}
\caption{Prototype BLE/Ultrasonic Protocol Message Timeline.}
\label{protocol}
\end{figure}



For the initial implementation we allocated up to 600 ms for receiver processing and 500 ms for the BLE data exchange. A single measurement between two phones would therefore take a total of 1.5 seconds, but this included a generous margin. Based on the discussion in the next section, this time could be reduced substantially.

\subsection{Alternative Protocols}

 Before settling on the above protocol, we considered several other approaches and discarded then. We discuss why we didn't pursue two of these promising alternatives here.  

\textbf{Receive and echo:} An initial design involved the first phone emitting a signal and the second phone receiving, amplifying, and ``echoing'' the signal back to the first phone. 

The benefit of the design is its simplicity; in particular, the task of the second phone is simple and all timing measurements are made with respect to time on the first phone, and the time difference measured at the first phone will include twice the time-of-flight and any loopback delays. 

Unfortunately, while this approach might work in clear outdoor environments, multipath reflections will also return to the first phone from objects in the environment. Distinguishing the desired return from the multipath reflection is challenging without significant processing. 

\textbf{Time difference of arrival between BLE and audio:} If a device were capable of sending a BLE packet and audio at the exact same time, any receiver would be able to learn the sender's distance by marking the time difference between when it received the start of the BLE packet and initially received the audio waveform, as in the Cricket system~\cite{cricket}. Unfortunately, the actual action of sending a packet (and, indeed, all of the RF capabilities of a smartphone), are controlled by the device's baseband, which runs an unmodifiable proprietary OS that is outside the purview of phone OS vendors. For example, the Qualcomm baseband on the Pixel 2 phone provides no explicit guarantees about the timing of packets, nor any feedback to the kernel software about when a packet was actually sent. 

\subsection{SonicPACT Ranging Protocol}
\label{sec:ranging}

A key challenge in estimating the range using time-of-flight between phones is the lack of accurate synchronization of time between the devices. At the speed of sound (346 meters/second at sea level), a timing offset of 10 ms is equivalent to a range error of 3.46 meters (11.4 feet). It is likely that independent phones will have varying offsets in practice, depending on variables such as the processing load, OS version, manufacturer Stock Keeping Unit (SKU), or other details, making it necessary to use an approach that can account for variability between phones. 

To account for these timing uncertainties, the protocol uses a \emph{loopback} measurement---each device estimates the time it takes to hear its own broadcasts in order to estimate its own processing delay. As long as this process takes place before the range between the phones has changed, and the timing offsets are the same between hearing its own signal and the signal from the other phone, the offsets will cancel and allow each phone to estimate the range. This is a well known technique for two-way time transfer, and is summarized by the steps below:
\begin{enumerate}
	\item Leader sends a waveform $x_1$ at $t_1 = \Delta_1+T_1+\delta_{T_1}$
		\begin{itemize}
			\item $\Delta_1$ is the Leader's offset from global time
			\item $T_1$ is the time the Leader thinks it is sending $x_1$
			\item $\delta_{T_1}$ is the delay before the Leader actually sends $x_1$
		\end{itemize}
	\item The Follower receives $x_1$ at $t_2 = \Delta_2+T_2-\delta_{R_2}$
		\begin{itemize}
			\item $\Delta_2$ is the Follower's offset from global time
			\item $T_2$ is the local time when the Follower detects $x_1$
			\item $\delta_{R_2}$ is the delay between $x_1$ being received and timestamped by software
		\end{itemize}
	\item The Follower sends a waveform $x_2$ at $t_3 = \Delta_2+T_3+\delta_{T_2}$
		\begin{itemize}
			\item $T_3$ is the time the Follower thinks it is sending $x_2$
			\item $\delta_{T_2}$ is the delay before the Follower actually sends $x_2$
		\end{itemize}
	\item The Leader receives $x_2$ at $t_4 = \Delta_1+T_4-\delta_{R_1}$
		\begin{itemize}
			\item $T_4$ is the local time when the Leader detects $x_2$
			\item $\delta_{R_1}$ is the delay between $x_2$ being received and timestamped by software
		\end{itemize}

\end{enumerate}

We now have two equations for the range between the phones:

\[R=c(t_2-t_1)=c(t_4-t_3)\]
where $c$ is the speed of propagation of sound. By adding the two equations (twice the range), substituting, and rearranging, the range is 
\begin{equation}
R=\frac{c}{2}[(T_4-T_1)-(\delta_{T_1}+\delta_{R_1})]-[(T_3-T_2)+(\delta_{T_2}+\delta_{R_2})],
\label{eq:orig}
\end{equation}
where the quantity within the first set of brackets is computed by the Leader and the quantity within the second set of brackets is computed by the Follower. This equation can also be obtained more directly from Figure~\ref{protocol}. Thus, each phone needs to know the time when it sends its own waveform, the time when it detects the waveform from the other phone, and its transmit/receive loopback delay $(\delta_T+\delta_R)$. If each phone also receives its own waveform to measure the loopback delay, say at $T_5$ on the Leader for the signal sent at $T_1$ and at $T_6$ on the Follower for the signal sent at $T_3$, then 

\[(\delta_{T_1}+\delta_{R_1}) = (T_5-T_1)\]
\[(\delta_{T_2}+\delta_{R_2}) = (T_6-T_3)\]
and substituting into the equation for range shows that the transmit times $T_1$ and $T_3$ also cancel. Therefore, the range can also be written as
\begin{equation}
R=\frac{c}{2}([T_4-T_5]+[T_2-T_6])
\label{eq:range}
\end{equation}
and each phone only needs to compute the difference in arrival times of the other phone’s waveform and their own waveform. In this case the variability in loopback delay is less of a concern. However, Eq. (\ref{eq:orig}) may still be useful for situations where reduced processing is necessary, since potentially the loopback delay is a constant calibration factor. Either way, after each phone has computed the quantity within the brackets, it exchanges its measurement with the other phone over BLE and both phones can estimate the range between them.

\subsection{Waveform Generation}
\label{sec:waveform}

In order for the ultrasonic ranging concept to be implemented, the smartphone's speaker and microphone hardware must be capable of passing signals at ultrasonic frequencies (ideally 20 kHz and above) without significant attenuation. Public documentation suggests that most newer iPhones support an audio sample rate of 48 KSPS, and so should have the digital bandwidth to represent signals up to 24 kHz. However, the analog hardware is likely not optimized for this range. As a first step, some simple experiments were done to evaluate the frequency response of the iPhones available to the group to get some idea of how well ultrasonic ranging would work.

\begin{figure}[]
\centerline{\includegraphics[width=\linewidth]{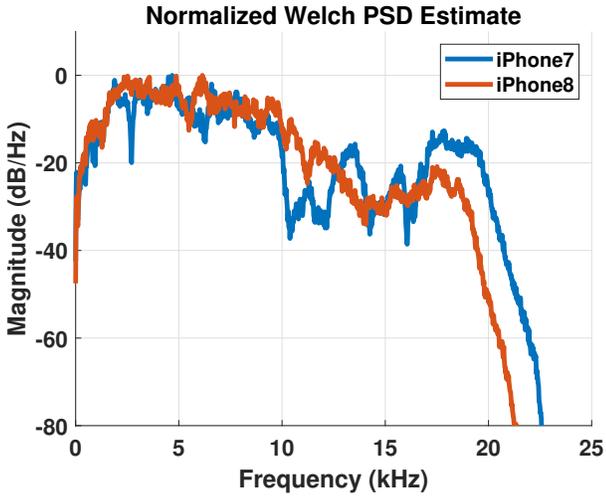}}
\caption{iPhone 7 and iPhone 8 Frequency Responses.}
\label{psd}
\end{figure}

We wrote an application to simultaneously play and record audio waveforms using Apple’s AVFoundation framework. The first test was to loopback White Gaussian Noise (WGN) from the speaker to microphone in order to characterize the frequency response. While this doesn’t inform us of the independent frequency responses of the speaker and microphone, it gives us a reasonable sense of the hardware capabilities. We used Welch’s method to estimate the Power Spectral Density (PSD) from the received samples. The results (Figure \ref{psd}) for an iPhone 7 and 8 show that both responses drop sharply above 20 kHz, but may be able to pass frequencies in the 17-19 kHz range. The attenuation profile is not ideal, but by using signals with a large time-bandwidth product we can expect a large processing gain and high-resolution time delay estimate.

Ranging systems, such as those used in RADAR, SONAR, and radio communications, typically use signals with a large time-bandwidth product \cite{richards}. Signals with a long duration can be integrated to boost signal-to-noise ratio (SNR) for detection, and signals with a wide bandwidth have fine time resolution (and therefore fine range resolution). Some popular signals include LFM chirps and Binary Phase Shift Keying (BPSK) modulated pulses. For the ultrasonic use case, we desired a signal that is:

\begin{itemize}
\item Uniquely identifiable for each phone
\item Simple to generate and detect
\item Capable of high-accuracy ranging
\end{itemize}

The Cramér-Rao Lower Bound (CRLB) is one starting point for understanding the minimum variance of an unbiased range estimator. For a signal in Additive White Gaussian Noise (AWGN), the CRLB for range is \cite{kay}:
\begin{equation}
\sigma^2_r \leq \frac{c^2}{\frac{E}{N_0/2}F^2},
\end{equation}
where $\sigma^2_r$ is the variance of the range, $c$ is the speed of propagation, $\frac{E}{N_0/2}$ is a SNR, and $F^2$ is the mean squared bandwidth of the signal. Note that the CRLB does not tell us how to estimate range, but it is informative to see that the variance is reduced linearly with the SNR and quadratically with the bandwidth of the signal.

\begin{figure}[]
\centerline{\includegraphics[width=\linewidth]{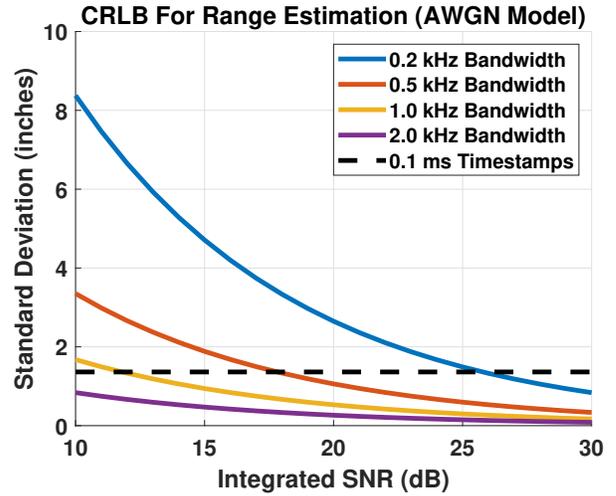}}
\caption{CRLB for the standard deviation of range error.}
\label{crlb}
\end{figure}

The CRLB for range is plotted in Figure \ref{crlb} as a function of integrated SNR for various rectangular (uniform) bandwidths. Also shown is an error floor if the device were to have variability in its reported timestamps with a standard deviation of 0.1 ms. It is worth noting that practical estimators require an integrated SNR of 13-16 dB for an unambiguous solution \cite{weinstein}. Based on this plot, we could choose a waveform with 200 Hz bandwidth and target an integrated SNR of 28 dB, or 500 Hz bandwidth and target an integrated SNR of 20 dB, and have a similar performance bound. 

The simple model used to derive the CRLB does not consider correlated noise or interference. We built a large margin into the signal design to help boost the SNR and overcome these effects while keeping the receiver processing relatively simple. As a starting point we used a waveform with 500 Hz bandwidth and 100 ms duration, which has a time-bandwidth product of 50 and can yield a processing gain of about 17 dB. Based on Figure \ref{psd}, the iPhone 8’s frequency response starts to significantly attenuate signals above 19 kHz, so we used a carrier frequency of 18.5 kHz. While this is not quite in the ultrasonic range, it should be inaudible to the majority of the population (particularly for adults).

We choose a modulation scheme such that each phone generates a waveform orthogonal to any other phone. Initially, we considered LFM chirps and BPSK-modulated signals, but pursued instead a noise-like waveform due to its ease of generation. The noise waveform has a high peak-to-average power ratio, and will therefore produce less output power than other modulations. Additional schemes should be considered for scenarios where phones may be in pockets, purses, or other sound-deadening material. We considered orthogonal chirps and chirp spread spectrum signals, but did not implement them.

\begin{figure}[]
\centerline{\includegraphics[width=\linewidth]{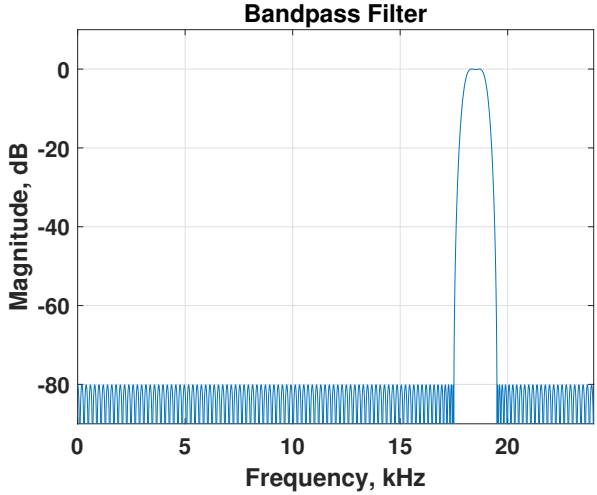}}
\caption{Bandpass filter centered at 18.5 kHz.}
\label{bpf}
\end{figure}

We generate WGN samples using a Gaussian Random Number Generator (RNG) seeded by a hash of a UUID. The WGN is then passed through a bandpass filter with 500 Hz of bandwidth centered at 18.5 kHz, as shown in Figure \ref{bpf}. Finally, a tone at 18.5 kHz is embedded at 10 dB below full scale. The reason for the tone is to aid in any frequency recovery that must be done by the receiver, since clock offsets or the Doppler effect will cause the received waveform to decorrelate with the reference copy. A Fast Fourier Transform (FFT) can be used to estimate the frequency of the received tone and enable the receiver to correct for any offset, avoiding the need to compute the full time/frequency ambiguity function.

\begin{figure}[]
\centerline{\includegraphics[width=\linewidth]{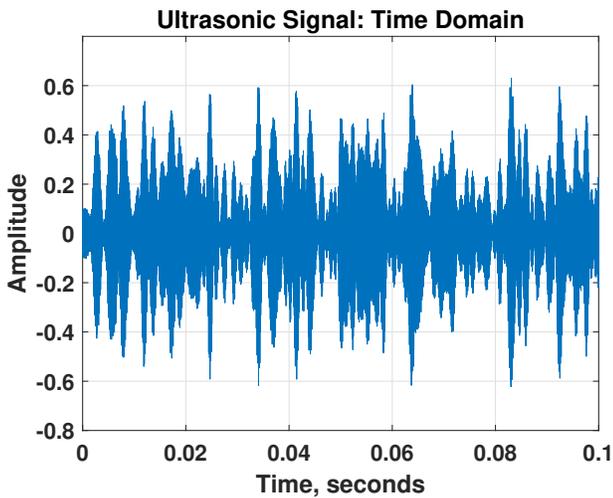}}
\caption{Ultrasonic signal in the time domain.}
\label{td}
\end{figure}

\begin{figure}[]
\centerline{\includegraphics[width=\linewidth]{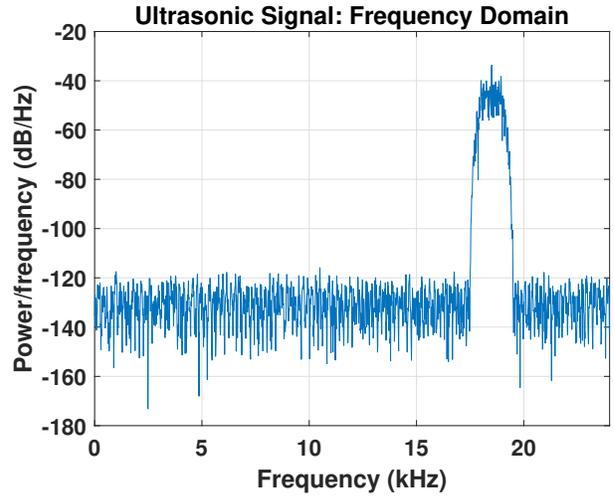}}
\caption{Ultrasonic signal in the frequency domain (note tone at 18.5 kHz).}
\label{fd}
\end{figure}

\begin{figure}[]
\centerline{\includegraphics[width=\linewidth]{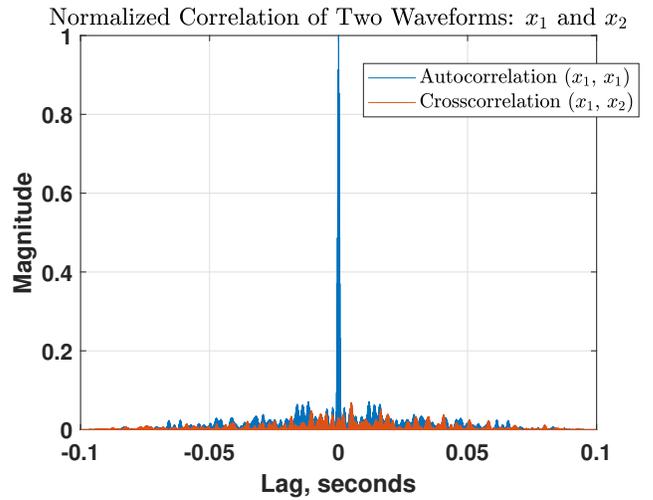}}
\caption{Correlation of two bandpass noise signals $x_1$ and $x_2$.}
\label{lag}
\end{figure}

\begin{figure}[]
\centerline{\includegraphics[width=\linewidth]{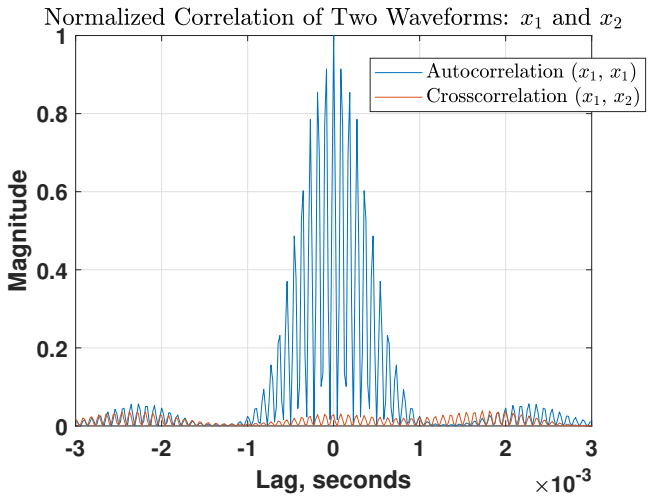}}
\caption{Expanded peak from \ref{lag} showing oscillations that result from directly processing a real bandpass signal.}
\label{lagzoom}
\end{figure}

An example signal is shown in the time and frequency domains in Figure \ref{td} and Figure \ref{fd}, respectively. The signal has a high correlation with itself when time aligned, low correlation with itself when not time aligned, and low correlation with signals generated using other UUIDs. These are favorable properties for multiple signals to coexist simultaneously on a channel with minimal interference to one another. An example of the correlation is shown in Figure \ref{lag} over the full span of the signal and in Figure \ref{lagzoom} expanded to +/- 3 ms around the peak. The oscillations are due to the fact that we are directly processing a real bandpass signal (without first shifting to baseband and filtering), however, with a large excess SNR we should be able to correctly detect the peak at the receiver.

Finally, to prevent audible clicks due to the discontinuities when the signal begins and ends, a linear ramp-up and ramp-down of 2.5 ms is applied to the signal before it is transmitted for iOS, and a 1.5 ms ramp is used for Android. These timings were determined empirically for the iPhone 7, iPhone 8, and Pixel 2 hardware specifically -- it is likely that other devices will require different ramps for the transmission to be inaudible.

A summary of the waveform design is shown in Table \ref{swfs}.

\begin{table}[]
\caption{Smartphone Waveform Parameters}
\begin{center}
\begin{minipage}{6cm}
\begin{tabular}{lllll}
\textbf{Parameter}        & \textbf{Value}           &  &  &  \\
Pulse Width               & 100 ms &  &  &  \\
Center Frequency          & 18.5 kHz                   &  &  &  \\
Sampling Rate             & 48 kHz                   &  &  &  \\
Bandwidth                 & 500 Hz                    &  &  &  \\
Modulation \footnote{Processing gain and orthogonality for robustness against co-channel interference}					& Bandpass WGN generated from BLE UUID	& & & \\
Rise/Fall Time \footnote{To reduce audible clicks}           & 2.5 ms  &  &  & 
\end{tabular}
\end{minipage}
\label{swfs}
\end{center}
\end{table}

\subsection{Waveform Detector}
\label{sec:detect}

\begin{figure}[]
\centerline{\includegraphics[width=\linewidth]{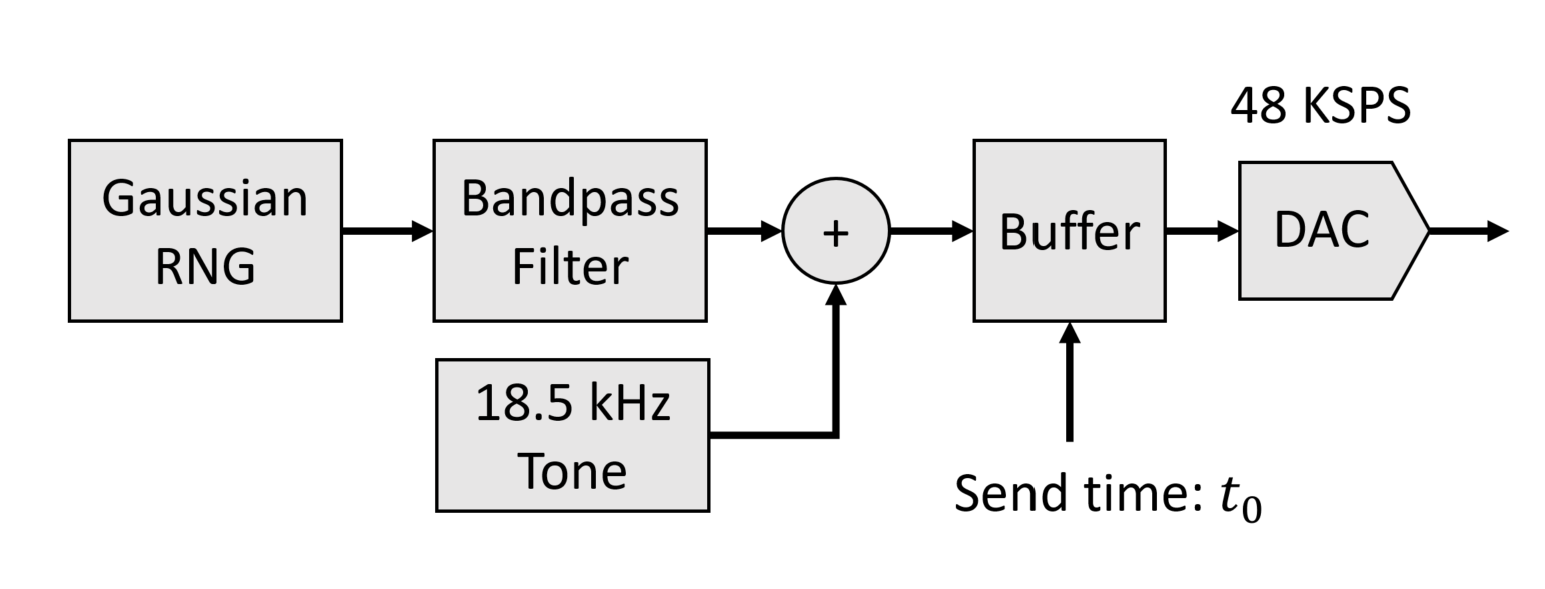}}
\caption{Transmitter processing block diagram.}
\label{det1}
\end{figure}

The detection processing is based on the concept of matched filtering, wherein we design a digital filter with its coefficients (taps) matched to a set of known waveform samples. As covered by the previous section, these samples are generated by a Gaussian RNG, filtered, and loaded into a buffer to be transmitted. This process is done twice: once based on a phone’s own UUID, and again for the UUID of the remote phone. A block diagram of the transmitter is shown in Figure \ref{det1}, where the contents of the buffer are transmitted at time $t_0$.

\begin{figure}[]
\centerline{\includegraphics[width=\linewidth]{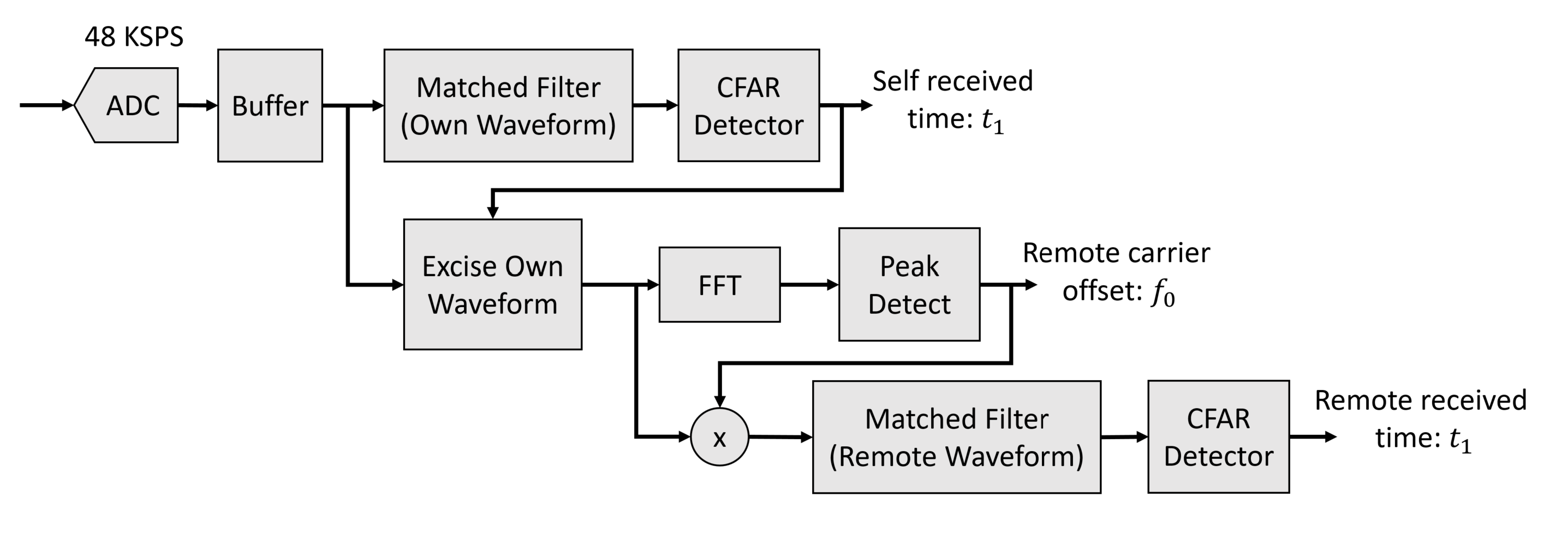}}
\caption{Receiver processing block diagram.}
\label{det2}
\end{figure}

The receiver was designed to perform batch processing based on a single buffer of samples returned with timestamp $t_1$. Based on experimenting with the iPhone 7 and 8, it was determined that a 400 ms buffer (19200 samples at 48 KSPS) could be reliably supported. Once a buffer was returned the first step was to apply a matched filter (correlate) with the phone’s own transmitted signal samples and perform peak detection to estimate the time of arrival as $t_2$. The time of arrival is then used for time domain excision (zeroing) of the phone’s own signal in the buffer, so that it will not interfere with the detection for the much weaker signal from the remote phone. This processing is shown by the top portion of Figure \ref{det2}.

After time domain excision the remaining samples contain the remote waveform along with any noise and interference. As mentioned in the previous section, any carrier frequency offset between the two phones (potentially due to motion and the Doppler effect) will cause the output of the matched filter to be reduced. Therefore, the next step is to perform frequency recovery by estimating an offset $f_0$ from the ideal carrier frequency of $f_c$. This is done by performing an FFT on the buffer (including samples that were zeroed out), finding the bin with the peak magnitude, and identifying the difference between that bin’s frequency and the nominal carrier frequency. The offset between the two is used to generate a sinusoid to be mixed with the buffer samples and correct for the offset. Note that the reason for including the zero samples in the FFT is to interpolate the frequency domain data (which helps in identifying the true peak).

\begin{figure}[]
\centerline{\includegraphics[width=\linewidth]{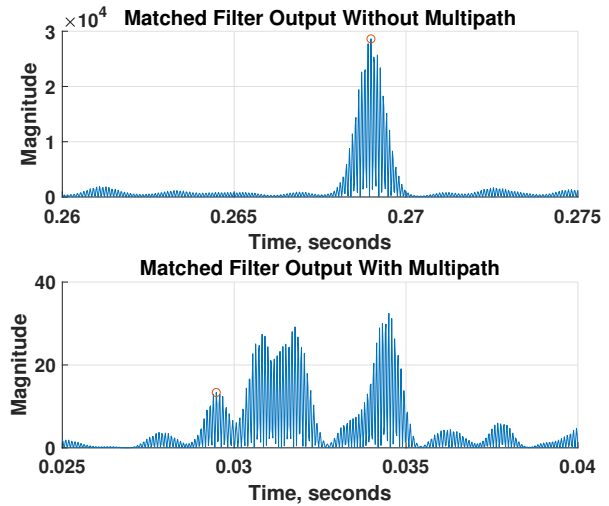}}
\caption{Matched filter output with and without multipath. The red circle in each figure is the true peak. The times on the $x$ axis of the two graphs are not comparable (i.e., time 0 is not when the signal was sent or received).}
\label{mf}
\end{figure}

After frequency correction the samples are processed by a matched filter designed for the remote phone’s transmitted signal. We used a basic detector with an adaptive threshold to attempt to find the first received copy of this signal by comparing the filter output power to an estimate of the background noise. An adaptive threshold is important for the processing to work when delayed copies of the signal arrive via multipath propagation, or when the background noise changes significantly due to interference. Some example matched filter outputs are shown in Figure \ref{mf}. The top plot shows a case without multipath and the two phones within inches of each other. The bottom plot shows multiple peaks from multipath off of the walls and furniture in a room and the phones 12 feet apart. Note the significant difference in magnitudes and potentially challenging problem detecting the first true peak in the bottom plot.

Ideally a large amount of data and truth would be collected to design a robust detector. However, the short time span of this project did not permit that, so a simple scheme was devised based on a limited amount of data. The adaptive threshold is set to 15 times the mean power output from the matched filter. The first sample out of the filter to cross this threshold is identified and used to return the next 50 samples (1 ms, about 1 foot in range). Within these samples we search for the two maxima and sort them by their times of arrival. If the maxima are separated by more than 15 samples (0.3 ms, about 3 inches in range), we choose the first maxima for the time of arrival estimate $t_3$. Otherwise we choose the largest of the two maxima for $t_3$. These heuristics seemed to work reasonably well on the initial data sets, but other schemes, such as a Constant False Alarm Rate (CFAR) detector \cite{richards}, may perform better.

Once the phone has estimated the times of arrival, it can compute the terms in Eq. (\ref{eq:range}) for use by the ranging algorithm. As described in \ref{sec:ranging}, the transmit time cancels and we are left with the difference in the receive times to exchange with the other phone.

\section{Implementation}
\label{sec:impl}

In this section, we describe system design aspects that are common to both iOS and Android implementations. We then go on to discuss OS-specific implementation details for iOS and Android. We discuss our experiments and results in the section that follows. All source code is available under MIT license at https://github.com/mit-ll/BluetoothProximity for iOS and  https://github.com/mspecter/SonicPACT for Android. 


The iOS app was written entirely in Swift, while the Android version is a combination of C++ and Java. For convenience and processing speed, both the iOS and Android implementations use libraries that take advantage of system hardware (e.g., ARM's NEON extensions). For iOS, we used GameplayKit \cite{gameplaykit} and vDSP \cite{vdsp}; for Android we used OpenCV \cite{opencv}.


%
%
%
%
%

\subsection{iOS Implementation Details}

The ranging algorithm and BLE/ultrasonic protocol were implemented on an iPhone 7 and 8 for initial evaluation. The implementation follows Figures \ref{det1}, \ref{det2}, and \ref{protocol}, apart from the FFT processing not being included in the real-time version due to time constraints (this was separately evaluated using offline processing and shown to perform as expected). This real-time limitation meant that tests could only be done between stationary phones, and that any motion or frequency offsets between the devices would result in degraded performance. Additionally, the implementation was only focused on the ultrasonic ranging measurements, so did not include the full PACT BLE protocol or contact duration estimator. Below are a few notes from working through the implementation:

\begin{itemize}
\item Sinusoid generation, such as that for the embedded tone, required double precision arithmetic for the phase argument. Using single precision resulted in audible tones during the transmission period (hypothesized to be harmonic distortion).
\item The Accelerate framework and vDSP provide highly efficient signal processing functions. As one data point, the convolution and correlation functions were over 2000 times faster using vDSP instead of nested for loops in Swift. 
\item The total receiver processing time for 400 ms worth of data was benchmarked to take 240 ms on the iPhone 7. This model is nearly 4 years old, and new ones are likely to require even less processing time. 
\item The iOS BLE stack does not provide a means for inserting custom data bytes into advertising packets, but it can encode an advertised name (string). Since we did not want to establish a full BLE connection to transfer data, we encoded our commands and data into the advertised name.
\item The delay between a BLE command being sent from one phone and received by the other appeared to nearly always be within the 50 ms window, but varied significantly. Even if we lost 10 ms of the waveform (due to the follower starting late), it is unlikely to be an issue because of the excess processing gain.
\end{itemize}

\subsection{Android Implementation Details}

We implemented and tested SonicPACT on a Google Pixel2. As with the iOS implementation, we left FFT processing and BLE contact duration estimation for future work.

Below are a few notes from working through the implementation:
\begin{itemize}
    \item Audio processing on Android is more complicated than on iOS. For example, initial implementations that attempted to record and play audio directly from Java resulted in unacceptable variations in timing and loss of specificity in time. We instead used Java only for the high-level protocol implementation, relying on C/C++ for most processing. We used Oboe \cite{Oboe}, a real-time audio processing library, for high-speed audio access.
    \item There are no OS-specific libraries for fast DSP processing (similar to vDSP on iOS). We instead had to rely on third party libraries for general matrix operations -- specifically OpenCV \cite{opencv} -- to provide equivalent processing speeds.
    \item Android's BLE stack is far more customizable than iOS, allowing for use of Bluetooth 5's extended advertisements to share timing and UUID information. Unfortunately, due to the privacy concerns associated with Bluetooth, this access also requires that the user accept that the app have access to ``fine-grained location information.''
    \item Unlike iOS, Android does not allow us to pre-buffer audio for a particular time delay. Instead, all audio is sent to the device through a high-priority process called MediaFlinger. MediaFlinger also performs OS-level mixing and filtering, further complicating matters.
\end{itemize}

Beyond the differences provided in the above list, Android is a fragmented ecosystem with a diverse set of hardware-dependent quirks. It is likely that optimizing SonicPACT for each device type will require some device-specific engineering.

%

\section{Evaluation}
\label{sec:eval}

\begin{figure}[t]
\includegraphics[width=\linewidth]{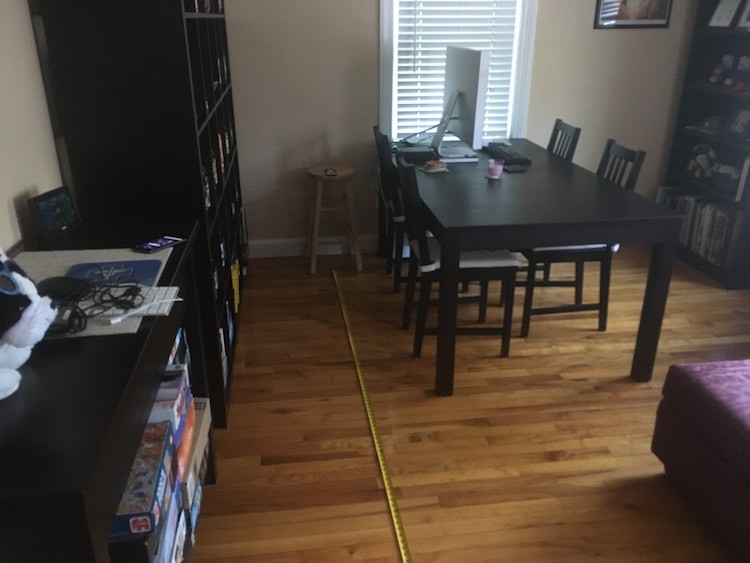}
\caption{Indoor test setup for iOS devices (one phone on stool at the end of the tape measure).}
\label{ios_indoor_setup}
\end{figure}

\begin{figure}[h]
\includegraphics[width=\linewidth]{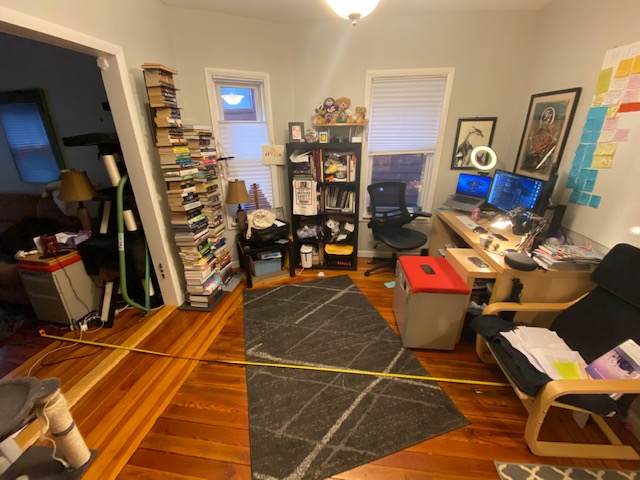}
\caption{Indoor test setup for Android devices.}
\label{android_indoor_setup}
\end{figure}



\begin{figure}[ht!]

\centering

\begin{subfigure}[b]{.5\textwidth}
\includegraphics[width=\textwidth]{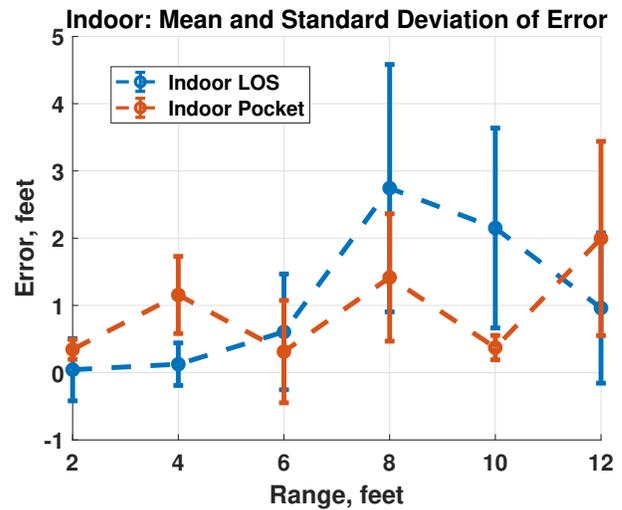}
\caption{Indoor range errors for iOS devices.}
\label{ios_indoor}
\end{subfigure}

\vspace*{.2in}

\begin{subfigure}[b]{.5\textwidth}
\centerline{\includegraphics[width=\linewidth]{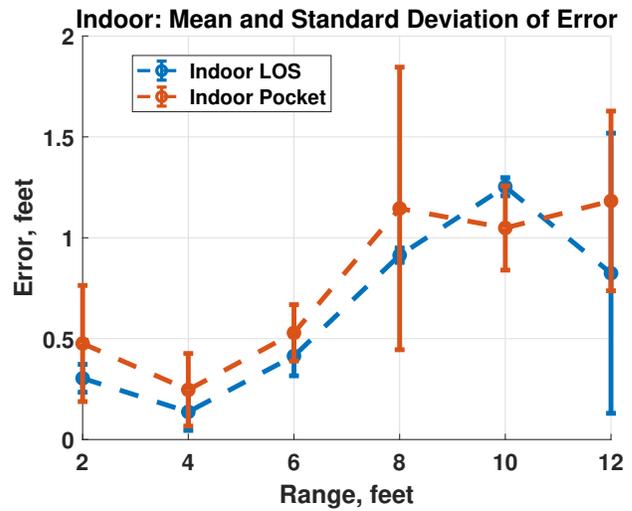}}
\caption{Indoor range errors for Android devices.}
\label{android_indoor_range}
\end{subfigure}

\caption{Indoor Results.}
\label{fig:iOS_Results}
\end{figure}

\begin{figure}[ht!]

\centering

\begin{subfigure}[b]{.5\textwidth}
\includegraphics[width=\textwidth]{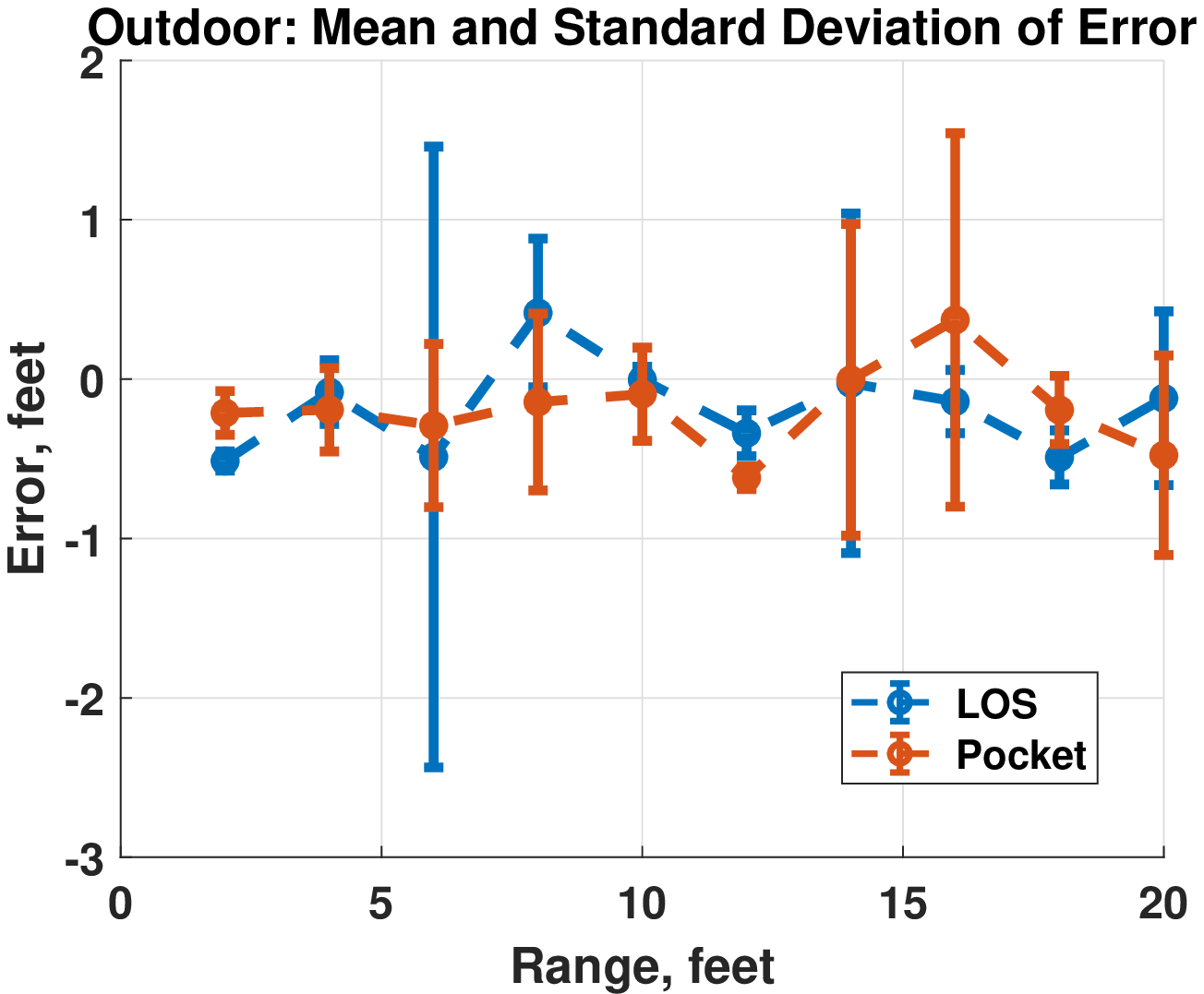}
\caption{Outdoor range errors for iOS devices.}
\label{ios_outdoor}
\end{subfigure}

\vspace*{.2in}

\begin{subfigure}[b]{.5\textwidth}
\centerline{\includegraphics[width=\linewidth]{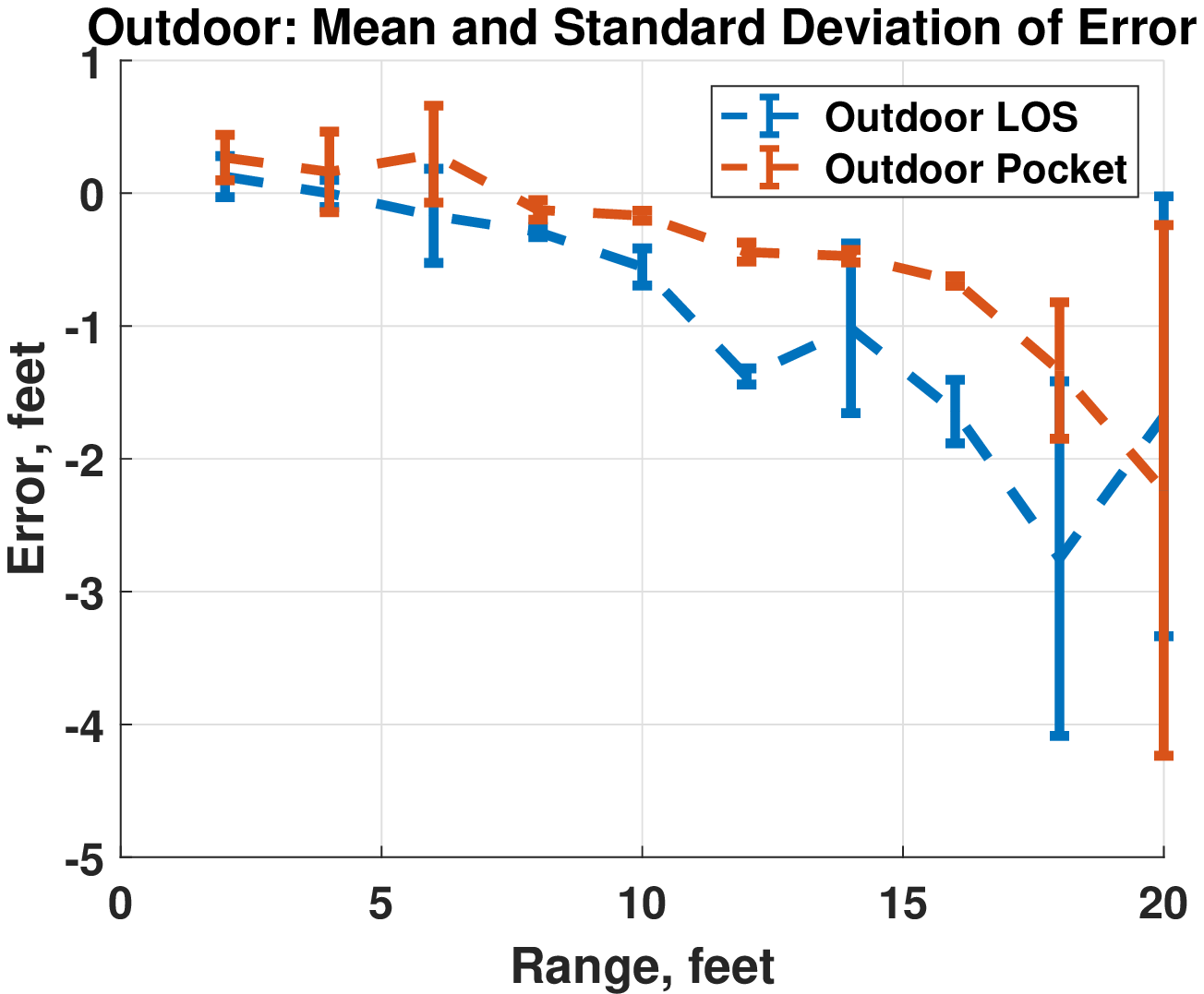}}
\caption{Outdoor range errors Android devices.}
\label{android_outdoor_range}
\end{subfigure}

\caption{Outdoor Results.}
\end{figure}

To evaluate our protocol, we took indoor and outdoor measurements to characterize the performance with various ranges between the phones in both Line-of-Sight (LOS) conditions and with one of the phones in a pocket. The test conditions were as follows:

\begin{itemize}
\item Outdoor, no obstructions, some multipath, ranges 2 to 20 feet (60 cm to 6 m) in 2-foot increments
\item Indoor, no obstructions, high multipath, ranges 2 to 12 feet (60 cm to 3.6 m) in 2-foot increments
\end{itemize}
In each case, we conducted two sets of tests:
	\begin{itemize}
		\item {\bf LOS:} One phone in hand, second phone on a chair
		\item {\bf Pocket:} One phone in hand, second phone in a pants pocket
	\end{itemize}

The indoor test setups for iOS and Android are shown in Figures \ref{ios_indoor_setup} and \ref{android_indoor_setup}, respectively. In each case we used a tape measure to record the true distance between the midpoints of each phone. We took at least ten measurements at each true distance. Both phones were set to one step below their maximum volume in the case of iOS, and at maximum volume in the case of Android.

\subsection{Metrics} 

We studied two sets of metrics. The first set is the {\bf miss rate} and the {\bf false alarm rate}, where we took the ranging data from each individual test and calculated the miss rate assuming a certain ``too close'' threshold. We show results for two such thresholds, 6 feet (1.8 m) and 8 feet (2.4 m). The miss rate measures how often the true distance is under the threshold but the measured distance was higher. The false alarm rate measures how often the measured distance was under the threshold but the true distance was higher than the threshold. Because the measurements have some noise, we permit a 1-foot slack; if the true distance is 6 feet and that is also the threshold, then we treat any measured distance $<$ 7 feet to be a correct detection (after all, if the measured distance were 6.3 feet, for example, any practical exposure notification system would consider that to be close enough to 6 feet). We use the same logic to assess false alarms.

We also report the fraction of measurements in each test where the measured distance was {\bf within 1 foot} (30 cm) of the truth.

Because the thresholds of interest for COVID-19 are in the 6-foot range, we restrict our analysis in the outdoor case to true distances $\leq$ 12 feet; larger distances artificially lower miss rates.

The second set of metrics evaluates the raw ranging performance showing the measured distance as a function of the true distance along with the standard deviations. It is worth noting that in many of these results single outliers cause the standard deviation to be large.

\subsection{Results}

\textbf{Indoor:} The miss rates and false alarm rates for the indoor tests are shown in Table~\ref{tab:missrates_indoors}.

\begin{table}[h]
    \centering
    \begin{tabular}{l|c|c|c|c|c}
     Test & 6-ft & 6-ft & 8-ft & 8-ft & Within \\
     & miss \% & false \% &  miss \% & false \% &  1 foot \%\\
     \hline
     LOS Android & 0.0 & 0.0 & 0.0 & 0.0 & 80.3\\
     LOS iOS & 5.1 & 0.0 & 11.9 & 0.0 & 57,6\\
     Pocket Android & 0.0 & 0.0 & 1.3 & 0.0 & 70.9 \\
     Pocket iOS & 3.3 & 0.0 & 13.3 & 0.0 & 56.7\\
    \end{tabular}
    \caption{Miss rates and false alarm rates for the indoor experiments. The ``Within 1 foot'' column shows the percentage of measured distances in the tests between 2 feet and 12 feet that were within 1 foot of the truth. These results show low miss rates and false alarm rates with a 1-foot slack across both platforms. We caution against drawing conclusions that Android is somehow better than iOS because the tests were done in different indoor conditions. It is likely that the iOS tests had higher multipath and Android tests had more sound absorbers.}
    \label{tab:missrates_indoors}
 \end{table}
 
The results of measured versus true distance in the indoor experiments are shown in Figure \ref{ios_indoor} 
for iOS and Figure \ref{android_indoor_range} 
for Android. For both charts, the points show the mean error and the vertical lines represent the standard deviation of the error. The errors are generally larger at shorter distances, especially at ranges beyond 6 feet. Based on a preliminary inspection, in our iOS experiments, strong multipath components are causing the detector to select a peak after the true peak, resulting in an estimated range a little greater than the true range.
The impact of phone placement (in hand or in a pocket) is small, possibly due to additional indoor multipath corrupting the SNR estimates. It is important to note that the iOS and Android experiments were done in different locations, so we caution against comparing them directly because the multipath effects are different.

\textbf{Outdoor:} The miss rates and false alarm rates for the outdoor tests are shown in Table~\ref{tab:missrates_outdoors}.

\begin{table}[h]
  
    \centering
    \begin{tabular}{l|c|c|c|c|c}
     Test & 6-ft & 6-ft & 8-ft & 8-ft & Within \\
     & miss \% & false \% &  miss \% & false \% &  1 foot \%\\
     \hline
     LOS Android & 0.0 & 0.0 & 0.0 & 0.0 & 82.9\\
     LOS iOS & 5.0 & 0.0 & 0.0 & 0.0 & 93.3\\
     Pocket Android & 0.0 & 0.0 & 1.3 & 0.0 & 100.0 \\
     Pocket iOS & 1.7 & 0.0 & 1.7 & 0.0 & 96.7\\
    \end{tabular}
    \caption{Miss rates and false alarm rates for the indoor experiments. The ``Within 1 foot'' column shows the percentage of measured distances in the tests between 2 feet and 12 feet that were within 1 foot of the truth. Outdoor performance is strong across both platforms both for LOS and when one device is in a pocket.}
    \label{tab:missrates_outdoors}
    
\end{table}

The outdoor results are shown in Figure \ref{ios_outdoor} 
for iOS and Figure \ref{android_outdoor_range} 
for Android.  Here we see that the mean error is within 1 foot of truth, and the standard deviation is contained within 1 foot for several ranges, particularly before the device is outside of the 10 ft radius.

For iOS, the large spread at ranges such as 6, 14, and 16 feet are from single outliers that can likely be eliminated with a more robust detector. We also see that the performance with one phone in and out of a pocket is similar. 

For Android, we note that the range errors 
worsen dramatically outside of 10 ft. We found experimentally that the Pixel2's speakers are forward-firing and highly directional, which likely explains some of the error rate, though there is likely some systematic error causing the negative bias. The TCFTL constraint is currently set (by public health authorities) at 6 ft, so this limitation is not a major problem.

\medskip

The results of all these experiments show that accurate ultrasonic ranging is possible on both iOS and Android smartphone platforms, and that SonicPACT is a viable technique for COVID-19 contact discovery and exposure notification. 




\section{Open Questions}

Though these early results are promising, several open questions remain and there are several opportunities for future work on this subject. It is also worth noting that the path to this demonstration was not a direct one; several naive approaches were considered and ultimately discarded. 

Our implementation serves as an existence proof that US ranging can work, but there are hurdles that must be overcome for practical use. Many of these can be considered for future work, but it is worth noting that some can only be resolved by trusted computing within the operating system.

\subsubsection{Operation at Scale} Though the use of a matched filter should minimize the impact of hidden terminals and other interference, careful protocol design will still be required to enable US ranging technology to operate effectively at scale. A crowded subway car with 50 devices and potential sources of interference is a significantly more challenging scenario than two devices in a quiet home or office. Additionally, US phenomenology is poorly understood and it is unclear how much interference at US frequencies will be present in real-world scenarios. The impact of multipath on measurement accuracy is unclear, as is the extent to which signal attenuation from fabric or body blockage will reduce the ability to make US measurements.

\subsubsection{Hardware Compatibility} The speakers and microphones used by various smartphone models have different frequency responses. The percentage of devices currently in-use by consumers that support reliable operation in the US frequency bands is unknown. However, anecdotal evidence observed by the PACT (and NOVID) teams suggests that this will not be a significant issue. Additionally, Lee et al. \cite{lee} have measured the speaker and microphone frequency responses of several devices, and they all appear to be quite similar. 

\subsubsection{Bluetooth Restrictions} iOS does not allow inserting a custom payload into a BLE advertising packet. However, it does allow setting the advertised name to a custom string. Because of this limitation, and the short time span of this project, the advertised name was encoded with commands and measurements for initial demonstration of the US ranging protocol. This approach was taken since it was preferred to avoid establishing a full BLE connection and only use advertising packets, similar to the Exposure Notification API. Android \emph{does} allow for this sort of extended advertising, but at the cost of requiring further location permissions from the operating system. We expect this can be solved on both platforms with greater OS integration.

\subsubsection{Audio Routing and Volume Control} On the iOS version of the application, we found that applications cannot control the precise route the audio will take. For instance, an attached pair of headphones (either analog or Bluetooth) will claim the audio interface and prevent the built-in speaker from being used by an app. On Android, the app is allowed to select the appropriate input/output devices, though there is significant variation on which devices exist, where they are, and which would be the most advantageous. Further, an app which overrides the user's volume settings may be problematic and interfere with normal smartphone usage unless properly designed.

\subsubsection{Efficiency \& Power} Outside of cursory analysis, we currently do not consider, nor do we measure, the impact on battery life. There is likely further engineering effort to ensure that power and efficiency are correct, but, given the complexity of this scheme as compared to normal activation commands, we believe that this protocol can be made to be efficient and robust, particularly if the existing voice-activation detection co-processors are used.

\subsubsection{Impact of US Broadcasts on Adults, Children, and Pets} Though adults typically cannot hear ultrasonic frequencies, they are often audible to children and pets. It is unclear to what extent the US measurement signals will be audible to children and/or pets and whether or not they will be bothersome or harmful. The PACT team has not, at this point, observed any such harmful effects. It would be useful consulting with an audiologist regarding the use of US frequencies for this purpose.

\subsubsection{Variability of the Speed of Sound} The speed of sound depends on the medium in which it is propagating, to include the temperature, pressure, and humidity. For a highly accurate implementation, the speed of sound may need to configured based on a coarse location (for example, the zip code) and/or a weather report.

\subsubsection{Variability of Devices} Android is a fragmented ecosystem with many devices running many different versions of the operating system, baseband, hardware profiles and SKUs. Timing is likely significantly different in these devices and, while this protocol does attempt to account for the majority of them, differences in mixing, hardware timing, and other variables leads us to believe that there is a strong possibility that there is a not-insignificant task of testing this protocol on a larger diversity of devices for particular quirks.

\section{Conclusion}
\label{sec:conclusion}
This paper provides a proof-of-concept that ultrasonic ranging is viable on commodity smartphones for the purposes of COVID-19 contact discovery and exposure notification, helping to reduce the false alarm and miss rates. A lower false positive rate may boost public confidence in the technique and encourage more widespread adoption. As noted in our discussion section, many unanswered questions remain, but these are surmountable by Apple and Google, as well as the research community. We urge Apple and Google to strongly consider ultrasonic ranging in general, and SonicPACT in particular, as a way to improve performance of the Exposure Notification APIs they have developed recently.


\bibliographystyle{IEEEtran}
\bibliography{refs.bib}

\end{document}